\documentclass[12pt,a4paper]{article}
\usepackage{amsmath,amsfonts,amssymb,graphicx}

\numberwithin{equation}{section}

\newcommand{\nc}{\newcommand}
\nc{\rnc}{\renewcommand}
\rnc{\title}[1]{{\Large\bf\mbox{}\\\medskip#1\bigskip\medskip\\}}
\rnc{\author}[1]{{\large #1\smallskip\\}}
\nc{\address}[1]{{\em #1\medskip\\}}
\def\gauss#1#2{\mbox{\small $\left[#1\atop #2\right]$}}
\def\gausst#1#2{\mbox{\small $\left[#1\atop #2\rule{0pt}{9pt}\right]$}}

\begin{document}
\begin{center}

\title{Lattice Approach to Excited TBA Boundary Flows:\\
Tricritical Ising Model}

\author{Giovanni Feverati\footnote{feverati@ms.unimelb.edu.au},
Paul A. Pearce\footnote{P.Pearce@ms.unimelb.edu.au}}

\address{Department of Mathematics and Statistics\\
University of Melbourne, Parkville, Victoria 3010, Australia}

\author{Francesco Ravanini\footnote{ravanini@bologna.infn.it}}

\address{INFN Sezione di Bologna, Dipartimento di Fisica\\
Via Irnerio 46, 40126 Bologna, Italy}

\end{center}

\setcounter{footnote}{0}

\begin{abstract}
\noindent We show how a lattice approach can be used to derive
Thermodynamic Bethe Ansatz (TBA) equations describing all excitations for
boundary flows. The method is illustrated for a prototypical flow of the
tricritical Ising model by considering the continuum scaling limit of the $
A_{4} $ lattice model with integrable boundaries. Fixing the bulk weights to
their critical values, the integrable boundary weights admit two boundary
fields
$\xi$ and
$\eta$ which play the role of the perturbing boundary fields $\varphi_{1,3}$
and $\varphi_{1,2}$ inducing the renormalization group flow between boundary
fixed points. The excitations are completely classified in terms of
$(\boldsymbol{m},\boldsymbol{n})$ systems  and quantum numbers but the
string content changes by certain mechanisms along the flow. For our
prototypical example, we identify these mechanisms and the induced map
between the relevant finitized Virasoro characters. We also solve the 
boundary TBA
equations numerically to determine the flows for the leading excitations.
\end{abstract}

\section{Introduction}

A problem of much current interest in Quantum Field Theory (QFT) is the
renormalization group (RG) flow between different boundary fixed points of a
Conformal Field Theory (CFT) that remains conformal in the bulk.
Many boundary flows have been studied including~\cite{btba,bdyFlows} 
Lee-Yang, minimal
models  and $c=1$ CFTs. In addition, approximate numerical scaling 
energies  can
be explored by use of the Truncated Conformal Space 
Approach (TCSA)~\cite{TCSA}.

In this letter, for boundary flows, we show how one can derive exact TBA 
equations~\cite{TBA} for all
excitations from a lattice approach~\cite{KlumpP,OPW,PearceN,PCA}. This
has not been possible in a field theory approach.
We illustrate
our approach in the context of a prototypical boundary flow of the Tricritical
Ising Model (TIM)~\cite{TIM,Affleck} with central charge $c=7/10$. This is an
interesting but relatively simple CFT with many applications in statistical
mechanics.  Following Cardy~\cite{Cardy}, the conformal boundary conditions
$B_{(r,s)}$ of the TIM are in one-to-one correspondence with the six primary
fields $\varphi_{(r,s)}\equiv\varphi_{(4-r,5-s)}$ labelled by the Kac labels
$(r,s)$ with $r=1,2,3$ and $s=1,2,3,4$. The Kac table of conformal weights
$\Delta_{r,s}$ is shown in Figure~\ref{kactable}. 
Of the many possible flows on a cylinder between boundary conditions 
$B_{(r,s)|(r',s')}$ (that is  $B_{(r,s)}$ on the left and  $B_{(r',s')}$ on 
the right) we only consider here flows involving a non-trivial boundary
$B_{(r,s)}$ on the left with a fixed trivial boundary $B_{(1,1)}$ on the 
right. For these conformal boundary conditions on the cylinder we use the 
special notation   
$$ {\cal B}_{(r,s)}=B_{(r,s)|(1,1)}. $$
In Figure~\ref{kactable} the possible integrable
flows~\cite{Affleck} between these conformal boundary conditions are 
shown schematically.

\begin{figure}[htbp]
\begin{center}
\begin{picture}(200,40)(0,0)
\put(0,0)
\mbox{\hspace{.5in}\Large
\begin{tabular}{r|c|c|c|l}
\multicolumn{5}{l}{$s$}\\
\cline{2-4}
2&$\ {1\over 10}\rule{0pt}{24pt}\ $&$\ {3\over 80}\ $&
$\ {3\over 5}\ $&\\[6pt]
\cline{2-4}
1&0&${7\over 16}\rule{0pt}{24pt}$&${3\over 2}$&\\[6pt]
\cline{2-4}
\end{tabular}}
\put(-154,-60){\Large
\begin{tabular}{rcccl}
&$\qquad\; 1\ $&$\ \;2\ $&$\ \;3\ $&$\ r\ $
\end{tabular}}
\put(-98,-2){\vector(0,-1){12}}
\put(-28,-2){\vector(0,-1){12}}
\put(-61,-2){\vector(0,-1){12}}
\put(-86,0){\vector(1,-1){17}}
\put(-36,0){\vector(-1,-1){17}}
\multiput(-74,-3)(-2,-2){8}{$\cdot$}
\multiput(-51,-3)(2,-2){8}{$\cdot$}
\put(-86,-14){\vector(-1,-1){3}}
\put(-36,-14){\vector(1,-1){3}}
\end{picture}
\end{center}
\vspace{.7in}
\caption{\label{kactable} \small Kac table of conformal weights 
$\Delta_{r,s}=((5r-4s)^2-1)/80$ of
the tricritical Ising model. We restrict to the lower half of the table with
$s=1,2$ using the usual identification $(r,s)\equiv(4-r,5-s)$. The integrable
flows between the six conformal boundary conditions ${\cal B}_{(r,s)}$ are 
shown schematically. The
$\varphi_{(1,3)}$ and $\varphi_{(1,2)}$ flows are indicated by solid and
dotted arrows respectively. We omit non-Cardy type boundary conditions.}
\end{figure}

In this letter we will focus on the prototypical flow obtained by 
perturbing the ${\cal B}_{(1,2)}$ boundary TIM by the relevant 
boundary operator $\varphi_{(1,3)}$
\begin{equation}
UV={\cal B}_{(1,2)}\mapsto {\cal B}_{(2,1)}=IR\label{protoflow},\qquad
\chi_{1,2}(q)\mapsto \chi_{2,1}(q)
\end{equation}
This flow induces a map between Virasoro characters $\chi_{r,s}(q)$ of the
theory where $q$ is the modular parameter. The physical direction of the flow
from the ultraviolet (UV) to the infrared (IR) is given by the relevant 
perturbations and is consistent with the
$g$-theorem~\cite{gthm} which asserts that the boundary entropy is 
reduced under
the flow
$g_{IR}<g_{UV}$. The boundary entropies are
\begin{equation}
g_{(r,s)}=\left({2\over 5}\right)^{1/4}
{\sin{\pi r\over 4}\,\sin{\pi s\over 5}\over
\sqrt{\sin{\pi\over 4}\,\sin{\pi\over 5}}},\qquad
g_{(1,1)}:g_{(2,1)}:g_{(1,2)};g_{(2,2)}=
1:\sqrt{2}:g:\sqrt{2}\,g
\label{g-fn}
\end{equation}
with $g=(1+\sqrt{5})/2$ the golden mean.

A more complete account of the full set of six
integrable flows of the TIM, including the second flow emanating from the 
${\cal B}_{(1,2)}$ boundary condition,  will be given in a subsequent
paper~\cite{FPRII}.  The approach outlined here, however, is quite
general and should apply, for example, to all integrable boundary flows of
minimal models~\cite{BPZ}.

\section{Scaling of Critical $A_4$ Model with Boundary Fields}

It is well known that the TIM is obtained as the
continuum scaling limit of the generalized hard square model of
Baxter~\cite{HardSq} on the Regime~III/IV critical line. This 
integrable lattice
model is the $A_4$ RSOS lattice model of Andrews-Baxter-Forrester~\cite{ABF}.
Specifically, in the presence of integrable boundaries, the scaling
energies of the TIM are obtained~\cite{OPW} from the scaling limit of the
eigenvalues of the commuting double-row transfer matrices
$\boldsymbol{D}(u)$~\cite{BPO} of the $A_4$ lattice model with $N$ faces in a
row. It has been shown by Behrend and Pearce~\cite{BP} that at 
criticality there
exist integrable boundary conditions associated with each conformal boundary
condition $(r,s)$ carrying 0,1 or 2 arbitrary boundary fields 
$\xi,\eta$. In the
present case the boundaries $(1,1)$ and $(3,1)$ 
admit no fields,
$(2,1)$ and $(1,2)$ admit a single field $\xi$ and 
$(2,2)$ admits two
boundary fields $\xi$ and $\eta$. These boundary fields induce the
$\varphi_{(1,3)}$ and $\varphi_{(1,2)}$ RG flows respectively so we 
identify the
thermal and magnetic boundary fields accordingly
\begin{equation}
\xi\sim\varphi_{(1,3)},\qquad \eta\sim\varphi_{(1,2)}.
\end{equation}

To consider boundary flows we must fix the bulk at its critical point and
vary the boundary fields $\xi,\eta$. But paradoxically, it has been
shown~\cite{OPW} that the fields $\xi,\eta$ are irrelevant in the sense
that they do not change the scaling energies if $\xi$ and $\eta$ are real
and restricted to appropriate intervals.
Explicitly, the cylinder partition functions
obtained~\cite{OPW,BP} from
$(r,s)$ integrable boundaries with $\xi,\eta$ real are independent of
$\xi,\eta$ and given by single Virasoro characters
\begin{equation}
{\cal B}_{(r,s)}=B_{(r,s)|(1,1)}: \qquad Z_{(r,s)|(1,1)}(q)=\chi _{r,s}(q).
\end{equation}
The reason for this is that in the lattice model the fields $\xi$ and $\eta$
control the location of zeros on the real axis in the complex plane of the
spectral parameter $u$ whereas only the zeros in the scaling regime, a
distance
$i\log N$ out from the real axis,  contribute in the scaling limit
$N\to\infty$. The solution is to scale the imaginary part of $\xi,\eta$ as
$\log N$. This is allowed because $\xi,\eta$ are arbitrary complex fields.

For our prototypical flow (\ref{protoflow}), following \cite{OPW,BP}, we
consider a cylindrical lattice with the $(1,1)$
boundary on the right (with no boundary field) and the $(2,1)$ boundary on
the left with boundary field $\xi_L$.  Explicitly, we now scale $\xi_L$ as
\begin{equation}
\label{scaling-xi}
\textrm{Re}(\xi_L)={\lambda\over 2}={\pi\over 5},\qquad
\textrm{Im}(\xi_L) = \frac{-\xi +\log N}{5}
\end{equation}
where $\xi$ is real.
In terms of boundary weights, the $(1,2)$ conformal boundary is reproduced in
the limit $\textrm{Im}(\xi_L)\rightarrow\pm\infty$ whereas the
$(2,1)$ conformal boundary is reproduced in the 
limit  $\textrm{Im}(\xi_L)\to 0$.

The scaling behaviour of the $A_4$ lattice model at the boundary fixed
points is described by the TBA equations of O'Brien, Pearce and
Warnaar~\cite{OPW} (OPW). In this letter, we generalize their analysis to
include the scaling boundary field~(\ref{scaling-xi}).
This parameter appears in the TBA equations only through the boundary
contribution $g(u)$ in (6.3) of OPW. For the flow (\ref{protoflow}), the
scaling limit of this function in the first analyticity strip
$-\frac{\lambda }{2}\leq
\textrm{Re}(u)\leq
\frac{3\lambda }{2}$ now becomes
\begin{equation}
\label{ghat}
\hat{g}_{1}(x,\xi )=\coth \frac{x+\xi }{2}
\end{equation}
with $\xi\rightarrow -\infty$ and $\xi\rightarrow +\infty$
corresponding to  ${\cal B}_{(1,2)}$ and ${\cal B}_{(2,1)}$ respectively.

\section{Classification and Flow of Excited States 
\mbox{${\cal B}_{(1,2)}\mapsto {\cal B}_{(2,1)}$}}

As in OPW, the scaling energies are classified by the patterns
of zeros of the eigenvalues of the double row transfer matrix
$\boldsymbol{D}(u)$. In the complex $u$ plane, they are organized
in two analyticity strips $-\frac{\lambda }{2}\leq
\textrm{Re}(u)\leq \frac{3\lambda }{2}$ and $ 2\lambda \leq
\textrm{Re}(u)\leq 4\lambda $.  For $N$ sufficiently large, there can be
only simple zeros in the middle of each strip (1-strings) or pairs of zeros
located  on the two borders of each strip, having the same imaginary part
(2-strings). Because of crossing symmetry, all the zeros appear in
complex conjugate pairs so the upper and lower half planes are related by
reflection in the real axis and this remains true for general values of
$\textrm{Im}(\xi_L)$.

The patterns of zeros are classified~\cite{OPW} by string content and
ordering.  Let $m_{1,2}\ge 0$ be the number of 1-strings in the upper half
plane in strips 1 and 2 and similarly $n_{1,2}\ge 0$ the number of 2-strings.
Then for the boundary fixed points ${\cal B}_{(1,2)}$ and ${\cal B}_{(2,1)}$, 
the string contents are classified by the $(\boldsymbol{m},\boldsymbol{n})$
systems
\begin{eqnarray}
UV={\cal B}_{(1,2)}\ (\mbox{$m_1$ odd, $m_2$ even}):&\!\!\!&
n_{1}=\frac{N\!+\!m_2\!+\!\sigma}{2}-m_1,
\:\  n_{2}=\frac{m_{1}-\sigma }{2}-m_{2}\quad\label{mn1} \\
IR={\cal B}_{(2,1)}\ \;(\mbox{$m_1$ odd, $m_2$ odd}):&\!\!\!&
n_{1}=\frac{N+m_2}{2}-m_1,
\qquad\! n_{2}=\frac{m_{1}+1}{2}-m_{2} \label{mn2}
\end{eqnarray}
where $N$ is odd. In the UV there are two distinct
$(\boldsymbol{m},\boldsymbol{n})$ systems labelled by $\sigma=\pm 1$. In the
scaling limit
$N\rightarrow \infty$, we have $n_1\sim N/2$ with $m_1, m_2, n_2=O(1)$.
For given string content, the relative ordering of 1- and 2- strings is
given by the quantum numbers $I_{k}^{(i)}$ where, for the
1-string $ y^{(i)}_{k} $, $I_{k}^{(i)}$ is the number of two-strings
whose location $ z^{(i)}_{l} $ is larger (further from the real axis) than
$y^{(i)}_{k}$.  Labelling the strings so that $ y^{(i)}_{k+1}>y^{(i)}_{k}$,
the quantum  numbers must satisfy
\begin{equation}
n_i\geq I_{1}^{(i)}\geq I_{2}^{(i)}\geq ...\geq
I_{m_{i}}^{(i)}\geq 0.
\end{equation}
In ${\cal B}_{(1,2)}$ with $\sigma=1$ there is a frozen 1-string, that is,
the 1-string furthest from the real axis in strip~1
is restricted to have no two-strings above it so that $I_{m_1}^{(1)}=0$.
With these quantum numbers, the states are uniquely labelled as
\begin{eqnarray}
UV={\cal B}_{(1,2)}: & \quad  & I=(I^{(1)}_{1}\, I^{(1)}_{2}\, 
...\,I^{(1)}_{m_1}\, |\,
I^{(2)}_{1}\, I^{(2)}_{2}\, ...\, I^{(2)}_{m_2}\, )_{\sigma } \\
IR={\cal B}_{(2,1)}: & \quad  & I=(I^{(1)}_{1}\, I^{(1)}_{2}\, 
...\,I^{(1)}_{m_1}\, |\,
I^{(2)}_{1}\, I^{(2)}_{2}\, ...\,  I^{(2)}_{m_2}\, )
\end{eqnarray}
In the scaling limit, $n_1$ is
infinite so in this limit the corresponding strip~1 quantum numbers are
unbounded.

Given the pattern of zeros, that is, the string content and quantum
numbers at the boundary fixed points, the conformal dimensions or scaling
energies of these states are uniquely determined~\cite{OPW} by
\begin{eqnarray}
\label{dim-12}
UV={\cal B}_{(1,2)}: \quad
E=\Delta_{(1,2)}\!+n^{UV}  &\!\!\!\!=\!\!\!&\frac{1}{10}+\frac{1}{4}
\boldsymbol{m} C \boldsymbol{ m}
-\frac{m_{1}-m_{2}}{2}\sigma +\sum _{i=1,2}\,
\sum _{k=1,..,m_{i}}\!\!\!\! I^{(i)}_{k}\quad
\\
\label{dim-21}
IR={\cal B}_{(2,1)}: \quad
E=\Delta_{(2,1)}\!+n^{IR} &\!\!\!\!=\!\!\!&
\frac{7}{16}-\frac{1}{2}+\frac{1}{4}
\boldsymbol{m} C \boldsymbol{m}
+\sum _{i=1,2}\, \sum _{k=1,..,m_{i}}\!\!\!\! I^{(i)}_{k}
\end{eqnarray}
where $\boldsymbol{m}=(m_1,m_2)$ and the Cartan matrix is
\begin{equation}
C= \left( \begin{matrix} 2 & -1 \\ -1 & 2 \end{matrix} \right)
\end{equation}
 From \cite{OPW}, we know that the above counting of states reproduces
exactly the expected UV and IR characters at the boundary fixed points but
there must be some mechanisms to change these classifications during the
flow.

\begin{figure}[t]
\hfill\includegraphics[width=0.3\linewidth]{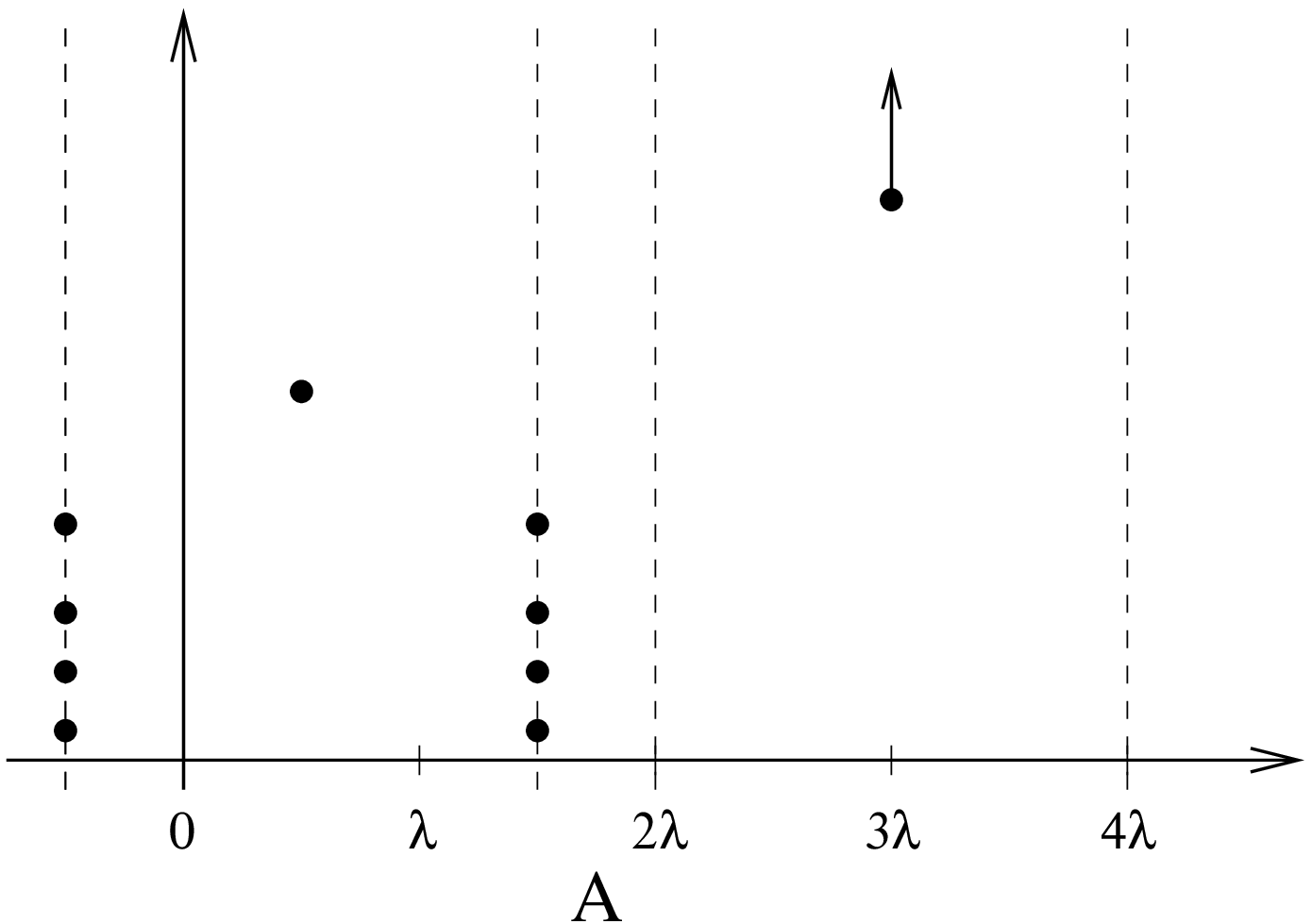}
\hfill\includegraphics[width=0.3\linewidth]{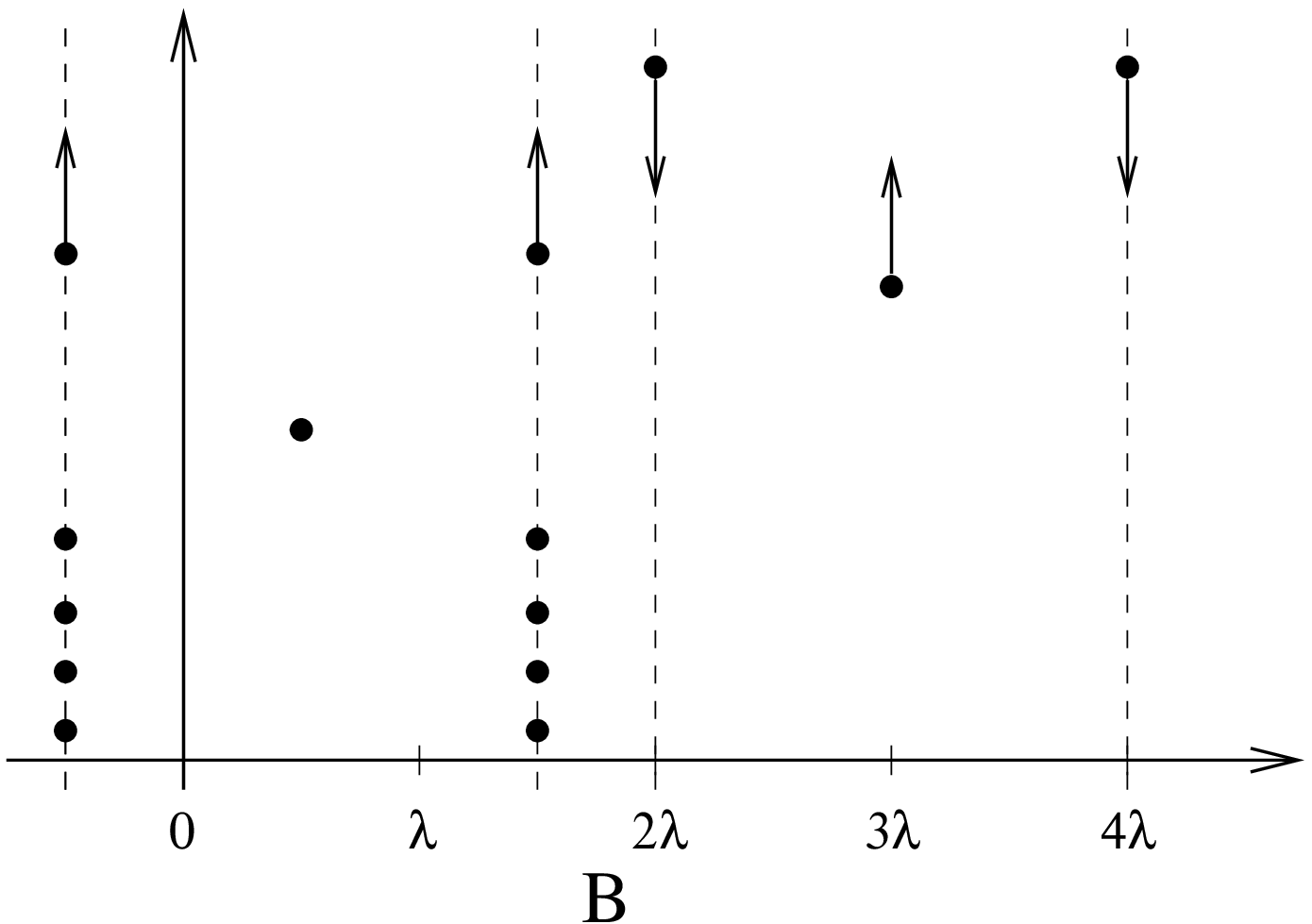}
\hfill\includegraphics[width=0.3\linewidth]{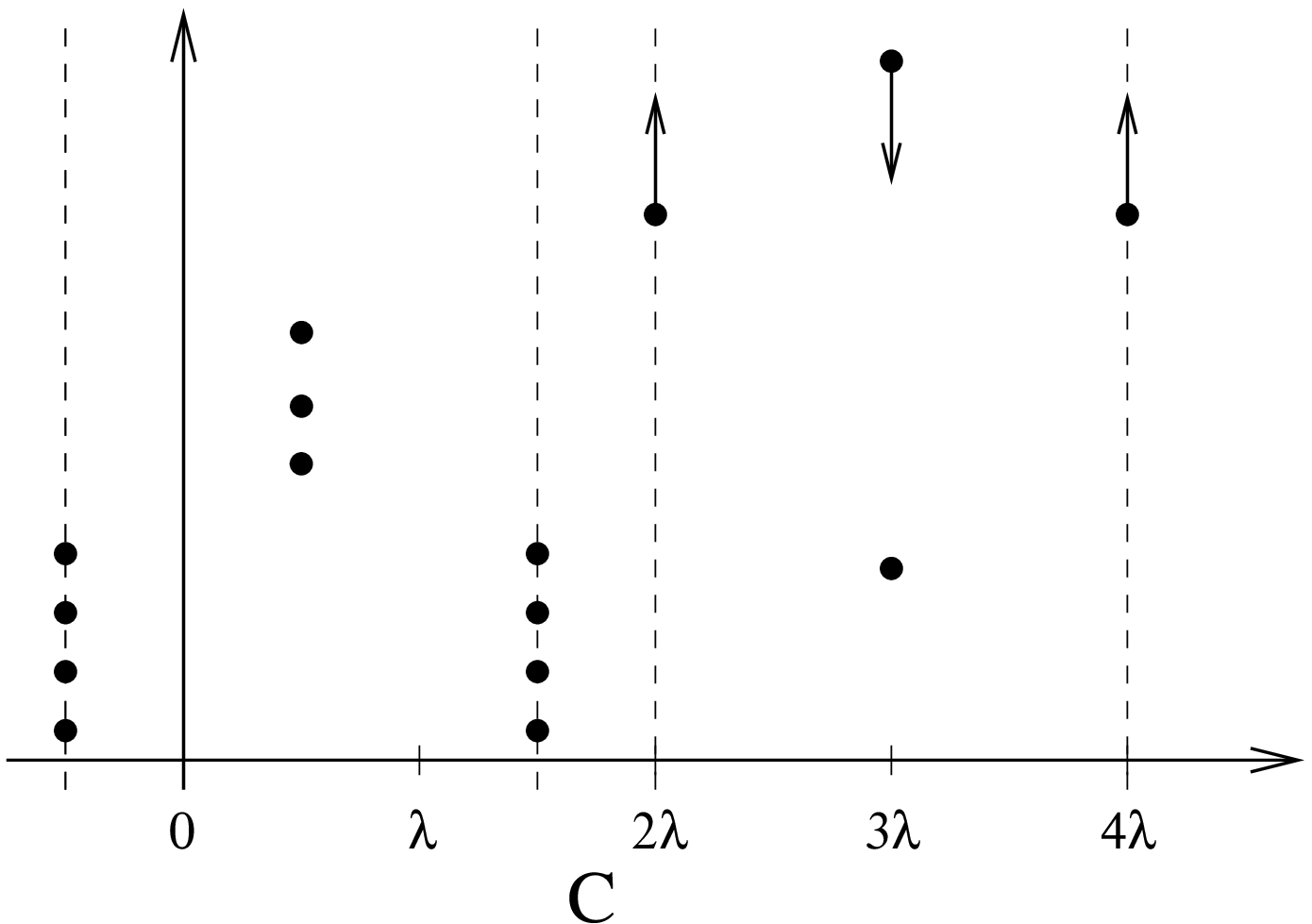}
\caption{\small \label{ABCmech} The three mechanisms A, B, C respectively
that change string content during the flow 
${\cal B}_{(2,1)} \mapsto {\cal B}_{(1,2)}$.
Note that this is the reverse of the physical flow.
These mechanisms are illustrated for the states: 
A, $(0|0)\,\mapsto\,(0)_{+}$;  
B, $(1|0)\,\mapsto\,(0)_{-}$;
C, $(0\,0\,0|1)\,\mapsto\,(0\,0\,0\,|0\,0)_{-}$.}
\end{figure}

\subsection{Three mechanisms for changing string content}

For convenience, in this section only, we consider the flow 
$IR={\cal B}_{(2,1)}\mapsto{\cal B}_{(1,2)}=UV$ 
which is the reverse of the physical flow (\ref{protoflow}). By
simple counting arguments, the eigenvalues of the double row transfer matrix
have $2N+4$ zeros in the periodicity strip for ${\cal B}_{(2,1)}$ 
but only $2N+2$
zeros for ${\cal B}_{(1,2)}$. Thus precisely two zeros, one in the upper half
plane and one in the lower half plane, must ``flow off to infinity" during
the flow. Indeed, from empirical observations based on direct numerical
diagonalization of small double row transfer matrices we find just three
mechanisms for changing the string content during the flow. These all
involve the farthest zeros (1 or 2 strings) from the real axis in the
upper half plane and are well supported by numerical analysis of the TBA
equations in Section~5.

The mechanisms for the flow $IR={\cal B}_{(2,1)}\mapsto
{\cal B}_{(1,2)}=UV$, as shown in Figure \ref{ABCmech}, are:
\begin{itemize}
\item[A.] The top 1-string in strip~2 flows to $+\infty$,
decoupling  from the system while $m_1$, $n_1$ and $n_2$ remain unchanged.
This mechanism applies in the IR when
$I^{(1)}_{m_1}=I^{(2)}_{m_2}=0$ and produces frozen states in the UV with
$\sigma=1$ and $I^{(1)}_{m_1}=0$.

\item[B.] The top 2-string in strip~1 and the top 1-string in
strip~2 flow to $+\infty$ and a 2-string comes in from $+\infty$ in
strip~2 becoming the top 2-string. Consequently, each $I^{(1)}_{j}$
decreases by
$1$ and each $I^{(2)}_{k}$ increases by $1$. This mechanism applies in the IR
when $I^{(1)}_{m_1}> 0$ and $I^{(2)}_{m_2}=0$ and produces states in 
the UV with
$\sigma=-1$ and either
$m_2=0$ or $I^{(2)}_{m_2}> 0$.

\item[C.] The top 2-string in strip~2 flows to $+\infty$ and a
1-string in strip~2 comes in from $+\infty$. Consequently,
each $I^{(2)}_{k}$ decreases by $1$.
This mechanism applies in the IR when $I^{(2)}_{m_2}> 0$ and produces states
in the UV with
$\sigma=-1$ and $I^{(2)}_{m_2}=0$.
\end{itemize}
Observe that these mappings are in fact one-to-one, that the 
1-strings in strip~1
are never involved in these mechanisms and that the change in parity 
of $m_2$ is
consistent with (\ref{mn1}) and (\ref{mn2})
\begin{equation}
A,B:\ \  m_2^{UV}=m_2^{IR}-1,\qquad
C:\ \  m_2^{UV}=m_2^{IR}+1
\label{m2change}
\end{equation}
The explicit mapping of the first 18 states is shown in Table~1.
The consistency and completeness of these three mechanisms is demonstrated by
the explicit mapping between the finitized characters associated with the UV
and IR fixed points. The situation here is very similar to the massless thermal
$\varphi_{1,3}$ flow~\cite{PCA} from the TIM to the critical Ising model and is
best described in terms of finitized Virasoro characters~\cite{FinChar}.

\begin{table}[htbp]
\begin{center}
\begin{tabular}{|c|r@{$\:\:\mapsto\:\:$}l|c||c|r@{$\:\:\mapsto\:\:$}l|c|}
\hline
$n$\rule[-4mm]{0mm}{10mm}& \multicolumn{3}{c||}{Mapping of states and 
mechanism} &
          $n$ & \multicolumn{3}{c|}{Mapping of states and mechanism} \\
\hline
$0$\rule[-1mm]{0mm}{6mm} & $(0)_{+}$ & $(0|0)$ & A &
          $5$ & $(4)_{-}$ & $(5|0)$ & B \\
$1$\rule[-1mm]{0mm}{6mm} & $(0)_{-}$ & $ (1|0)$ & B &
          $5$ & $(1\,1\,0)_{+}$ & $(1\,1\,0|0)$ & A \\
$2$\rule[-1mm]{0mm}{6mm} & $(1)_{-}$ & $(2|0)$ & B &
          $5$ & $(2\,0\,0)_{+}$ & $(2\,0\,0|0)$ & A \\
$3$\rule[-1mm]{0mm}{6mm} & $(2)_{-}$ & $(3|0)$ & B &
          $6$ & $(1\,1\,0|0\,0)_{-}$ & $(1\,1\,0|1)$ & C \\
$3$\rule[-1mm]{0mm}{6mm} & $(0\,0\,0)_{+}$ & $(0\,0\,0|0)$ & A &
          $6$ & $(2\,0\,0|0\,0)_{-}$ & $(2\,0\,0|1)$ & C \\
$4$\rule[-1mm]{0mm}{6mm} & $(0\,0\,0|0\,0)_{-}$ & $(0\,0\,0|1)$ & C &
          $6$ & $(5)_{-}$ & $(6|0)$ & B \\
$4$\rule[-1mm]{0mm}{6mm} & $(3)_{-}$ & $(4|0)$ & B &
          $6$ & $(0\,0\,0)_{-} $ & $(1\,1\,1|0)$ & B \\
$4$\rule[-1mm]{0mm}{7mm} & $(1\,0\,0)_{+}$ & $(1\,0\,0|0)$ & A &
          $6$ & $(2\,1\,0)_{+}$ & $(2\,1\,0|0)$ & A \\
$5$\rule[-1mm]{0mm}{6mm} & $(1\,0\,0|0\,0)_{-} $ & $(1\,0\,0|1)$ & C &
          $6$ & $(3\,0\,0)_{+}$ & $(3\,0\,0|0)$ & A \\
\hline
\end{tabular}
\end{center}
\caption{\label{table:states} \small Explicit mapping of states in the order 
of increasing energy for the first 18
energy levels, where $n=n^{UV}=n^{IR}$ is the excitation level as in 
(\ref{dim-12})
and (\ref{dim-21}). For some higher levels $n^{UV}\neq n^{IR}$.}
\end{table}

\subsection{RG mapping between finitized characters}
Finitized characters are the generating functions for the
scaling energies of the $A_4$ model for finite $M\times N$ lattices:
\begin{equation}
\chi_{r,s}^{(N)}(q)= q^{-c/24}\sum_E q^E=q^{-c/24}\; \textrm{Tr}
\left({\boldsymbol{D}^{(N)}(u)\over D^{(N)}_0(u)}\right)^{M/2}
\end{equation}
where the modular parameter is $q=\exp{(-\pi \sin (5u)M/N)}$ and
$D^{(N)}_0(u)$ is the largest eigenvalue of the double-row
transfer matrix $\mathbf{D}^{(N)}(u)$ with $(r,s)$ boundary conditions.
The finitized characters involve Gaussian polynomials which satisfy a recursion
used repeatedly below
\begin{equation}
\gauss{m+n}{m}=
\sum_{I_1=0}^n \sum_{I_2=0}^{I_1}\cdots \sum_{I_m=0}^{I_{m-1}} 
q^{I_1+\ldots+I_m},
\qquad
\gauss{n}{m}=\gauss{n-1}{m-1} + q^m \gauss{n-1}{m}
\end{equation}

Let us return to the map induced by the physical flow 
$UV={\cal B}_{(1,2)}\mapsto {\cal B}_{(2,1)}=IR$.  
We can write the UV finitized character as
\begin{eqnarray}
\chi_{(1,2)}^{(N)}(q)&\!\!\!=\!\!\!&\displaystyle 
q^{-\frac{c}{24}+\frac{1}{10}}
\!\!
\sum_{\sigma,m_1,m_2^{UV}}q^{\frac{1}{4}\boldsymbol{m}^{UV}C\boldsymbol{m}^{UV}}
q^{-\frac{\sigma}{2}(m_1-m_2^{UV})}
\gausst{m_1+n_1^{UV}-\delta_{\sigma,1}}{m_1-\delta_{\sigma,1}}
\gausst{m_2^{UV}+n_2^{UV}}{m_2^{UV}}
\nonumber\\
&\!\!\!=\!\!\!&q^{-\frac{c}{24}+\frac{1}{10}} \!\!
\sum_{m_1,m_2^{UV}}q^{\frac{1}{4}\boldsymbol{m}^{UV}C\boldsymbol{m}^{UV}}
\left\{q^{-\frac{1}{2}(m_1-m_2^{UV})}\gauss{{N+m_2^{UV}-1\over 2}}{m_1-1}
\gausst{{m_1-1\over 2}}{m_2^{UV}}\right. \quad\\
&&\hspace{-0.0in}\mbox{}+\left.q^{\frac{1}{2}(m_1-m_2^{UV})}
\gauss{{N+m_2^{UV}-1\over 2}}{\phantom{H}m_1^{\phantom{H}}}\left(
q^{m_2^{UV}}\gausst{{m_1-1\over 2}}{m_2^{UV}}+
\gausst{{m_1-1\over 2}}{m_2^{UV}-1}\right)\right\}\nonumber
\end{eqnarray}
where the three terms correspond to the three mechanisms A, B, C respectively.
This is confirmed by simple counting in the limit $q\to 1$ when the Gaussian
polynomials reduce to binomials. Here $m_1=m_1^{UV}=m_1^{IR}$ does not change
under the flow.
Assuming (\ref{dim-12}), (\ref{dim-21}) and mechanisms A, B, C hold for all $N$
we can uniquely map, $q^{E^{UV}}\mapsto q^{E^{IR}}$, energy level by energy
level.
In effect, this means we replace $m_2^{UV}$ with $m_2^{IR}$ using
(\ref{m2change}) and the energies at the base of each Gaussian tower change
according to
\begin{eqnarray}
\mbox{A:}&&q^{\frac{1}{4} \boldsymbol {m}^{UV} C \boldsymbol {m}^{UV}
- \frac{1}{2}(m_1-m_2^{UV})} \mapsto
q^{\frac{1}{4} \boldsymbol {m}^{IR} C \boldsymbol {m}^{IR}-\frac{1}{2}} \\
\mbox{B:}&&
q^{\frac{1}{4} \boldsymbol{m}^{UV} C \boldsymbol{m}^{UV}
+ \frac{1}{2}(m_1+m_2^{UV})} \mapsto
q^{\frac{1}{4}  \boldsymbol {m}^{IR} C \boldsymbol {m}^{IR}-\frac{1}{2}}
q^{m_1} \\
\mbox{C:}&&
q^{\frac{1}{4}  \boldsymbol {m}^{UV} C \boldsymbol {m}^{UV}
+ \frac{1}{2}(m_1-m_2^{UV})} \mapsto
q^{\frac{1}{4} \boldsymbol {m}^{IR} C \boldsymbol {m}^{IR}-\frac{1}{2}}
q^{m_2^{IR}}
\end{eqnarray}
Putting all this together we obtain the desired mapping between finitized
characters
\begin{eqnarray}
\chi_{(1,2)}^{(N)}(q)&\mapsto&
\displaystyle q^{-\frac{c}{24}+\frac{7}{16}-\frac{1}{2}}
\sum_{m_1,m_2^{IR}} q^{\frac{1}{4}\boldsymbol{m}^{IR}C\boldsymbol{m}^{IR}}
\left\{\gauss{{N+m_2^{IR}-2\over 2}}{m_1-1\rule{0pt}{7pt}}
\gausst{{m_1-1\over 2}}{m_2^{IR}-1}\right.
\nonumber\\
&&\hspace{0.25in}\mbox{}+q^{m_1}
\gauss{{N+m_2^{IR}-2\over 2}}{m_1\rule{0pt}{7pt}}
\gausst{{m_1-1\over 2}}{m_2^{IR}-1}
\left. + q^{m_2^{IR}}
\gauss{{N+m_2^{IR}\over 2}}{m_1\rule{0pt}{7pt}}
\gausst{{m_1-1\over 2}}{m_2^{IR}}\right\} \\[6pt]
&=&\displaystyle q^{-\frac{c}{24}+\frac{7}{16}-\frac{1}{2}}
\sum_{m_1,m_2^{IR}} q^{\frac{1}{4}\boldsymbol{m}^{IR}C\boldsymbol{m}^{IR}}
\gauss{{N+m_2^{IR}\over 2}}{m_1\rule{0pt}{7pt}} \left(
\gausst{{m_1-1\over 2}}{m_2^{IR}-1}
+ q^{m_2^{IR}}
\gausst{{m_1-1\over 2}}{m_2^{IR}}\right) \nonumber\\[6pt]
&=& \displaystyle q^{-\frac{c}{24}+\frac{7}{16}-\frac{1}{2}}
\sum_{m_1,m_2^{IR}} q^{\frac{1}{4}\boldsymbol{m}^{IR}C\boldsymbol{m}^{IR}}
\gauss{{N+m_2^{IR}\over 2}}{m_1\rule{0pt}{7pt}}
\gausst{{m_1+1\over 2}}{m_2^{IR}}
\;=\; \chi_{(2,1)}^{(N)}(q)  \nonumber
\end{eqnarray}

\section{Boundary TBA Equations ${\cal B}_{(1,2)}\mapsto {\cal B}_{(2,1)}$}
The derivation of TBA equations is a straightforward extension of OPW 
to include
the $\xi$-dependent boundary term (\ref{ghat}).
The TBA equations are
\begin{eqnarray}
\epsilon_1 (x) & = & -\log \hat{g}_{1}(x,\xi)-
\sum ^{m_{1}}_{j=1}\log \tanh (\frac{y_{j}^{(1)}-x}{2})-
K*\log (1-e^{-\epsilon_2(x)})  \label{TBA-1} \\
\epsilon_2(x) & = & 4e^{-x}-
\sum ^{m_{2}}_{k=1}\log \tanh (\frac{y_{k}^{(2)}-x}{2})-
K*\log (1-e^{-\epsilon_1(x)}) \label{TBA-2}
\end{eqnarray}
where $*$ denotes convolution, the kernel is
$K(x)=1/ (2\pi \cosh \,x)$
and the auxiliary variables $y^{(i)}_{k}$ denote the scaled location of the
1-strings in strip~$i$.
Similarly, the scaling energies $E(\xi)$ are
\begin{equation}
\label{c-tilde}
\displaystyle E(\xi)-\frac{c}{24}=\frac{1}{\pi }
\lim_{R\rightarrow 0} RE(R) =
\displaystyle \frac{1}{\pi }\left[ \sum ^{m_{1}}_{j=1}2e^{-y^{(1)}_{j}}-
\int ^{\infty }_{-\infty }\! \! \frac{dx}{\pi }e^{-x}
\log (1-e^{-\epsilon_{2}(x)}) \right]
\end{equation}
where the limit $R\rightarrow 0$ means that all these computations
are performed at the critical temperature of the TIM.
For the lowest level $E(\xi)$ interpolates between the conformal weights
$\Delta=1/10$ and $7/16$ at the boundary fixed points.

As in OPW the auxiliary variables $y^{(i)}_{k}$ are determined by a set of
auxiliary equations determining the locations of the 1-strings
\begin{eqnarray}
\epsilon_{2}(y^{(1)}_{j}-i\frac{\pi }{2}) & = & n_j^{(1)} i \pi,
\qquad \textrm{1-strings in strip~1}\label{location-1} \\
\epsilon_{1}(y^{(2)}_{k}-i\frac{\pi }{2}) & = & n_k^{(2)} i \pi,
\qquad \textrm{1-strings in strip~2}\label{location-2}
\end{eqnarray}
where $n_k^{(i)}$ are given in the IR by the \emph{quantization conditions}
\begin{eqnarray}
\label{nk1-21}
n_{j}^{(1)} & = & 2(m_{1}-j+I_{j}^{(1)})+1-m_{2},\quad j=1,2,\ldots,m_1\\
\label{nk2-21}
n_{k}^{(2)} & = & 2(m_{2}-k+I_{k}^{(2)})+1-m_{1},\quad k=1,2,\ldots,m_2
\end{eqnarray}
The integers $n_k^{(i)}$ can change during the flow due to winding of 
phases. The
appropriate form in the UV is given in OPW. Observe that the boundary 
field $\xi$
appears explicitly only in the function
$\hat{g}_1$.

The 2-strings do not appear in the TBA equations.
However, they are involved in mechanisms B and C.
To follow the movement of 2-strings along the flow we note that their locations
$z^{(i)}_{k}$ satisfy the same equations (\ref{location-1}), (\ref{location-2})
with the substitution
$y^{(i)}_{k}\rightarrow z^{(i)}_{k}$
and the appropriate choice of the quantum numbers.
We note that in general, and in particular for the TIM, it remains an 
open problem
in this approach to obtain the  $g$-function flow~\cite{btba,g-flow} that 
interpolates
between the boundary entropies (\ref{g-fn}) at the fixed points.

\newpage

\setlength{\unitlength}{1.0mm}
\begin{figure}[h]
\begin{center}
\begin{picture}(148,201)
\put(8,1){$\chi_{1,2}(q)$} \put(75,1){$\xi$} 
\put(131,1){$\chi_{2,1}(q)$}
\put(0,187){$E(\xi)$}
\put(125,27){1}\put(125,54){1}\put(125,81){1}
\put(125,108){2}\put(125,135){3}\put(125,162){4}\put(125,189){6}
\put(25,19){1}\put(25,46){1}\put(25,73){1}
\put(25,100){2}\put(25,127){3}\put(25,156){4}\put(25,181){6}
\put(5,5){\includegraphics[width=0.85\linewidth]{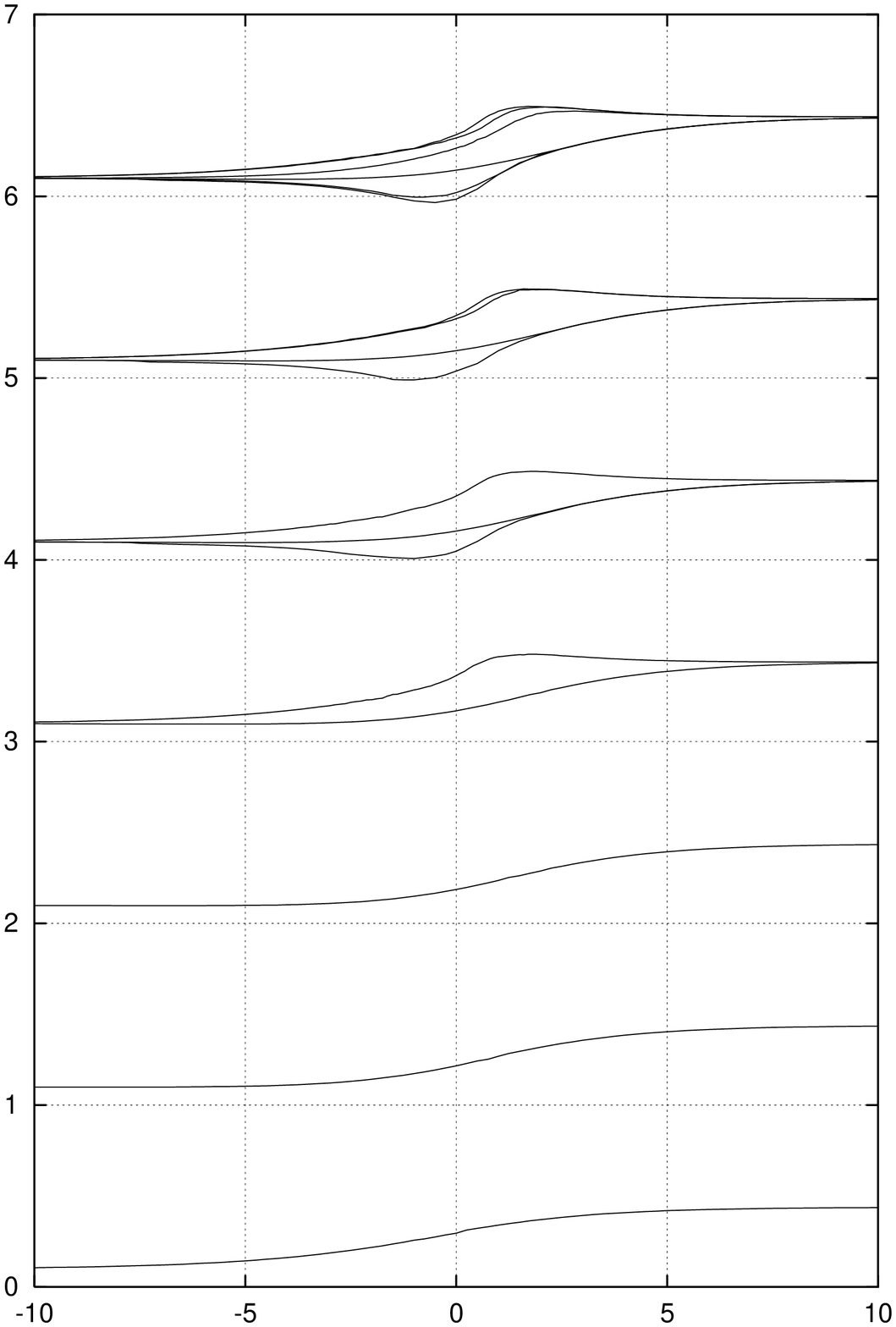} }
\end{picture}  
\end{center}
\caption{\small \label{figure:energies}  The flow of scaling energies
$E(\xi)$ between the ${\cal B}_{(1,2)}$ and ${\cal B}_{(2,1)}$ 
boundary fixed points 
as obtained by the numerical solution of the TBA equations. 
Only scaling energies with $\Delta<7$ are shown as in Table~1. 
The indicated degeneracies in the UV (left) and IR (right) are 
consistent with the degeneracies of the Virasoro characters
$\chi_{1,2}(q)=\chi_{1/10}(q)$ and 
$\chi_{2,1}(q)=\chi_{7/16}(q)$ respectively.}
\end{figure}

\clearpage

\section{Numerical Analysis of the RG Flow 
${\cal B}_{(1,2)}\mapsto {\cal B}_{(2,1)}$}

The TBA equations of the previous section can be solved
numerically by an iterative
procedure but there are some subtleties. The process starts with an inital
guess, for the pseudoenergies
$\epsilon_i(x)$ and 1-string locations, close to the UV or IR fixed points. The
flow is followed by progressively incrementing or decrementing the 
boundary field
$\xi$. At each value of $\xi$, the TBA equations are used to update the
pseudoenergies $\epsilon_i(x)$ and then these are used in the 
auxiliary equations
to update the locations of the 1-strings, and so on, until a stable solution is
reached.  In our preliminary numerical study we have considered all the states
with $\Delta<7$  as shown in  Figure~\ref{figure:energies}. The corresponding
mappings of UV and IR quantum numbers and the mechanisms responsible are listed
in Table~\ref{table:states}. These flows, and in particular the 
behaviour of the
1- and 2-strings, confirm the mechanisms A, B, C.

The most important new feature of the numerics is the existence, due 
to the term
(\ref{ghat}), of a pole at $x=-\xi$ which marches through the center of the
analyticity strip~1 as $\xi$ is varied. This leads to some inherent numerical
instabilities as the pole successively encounters the 1-string locations in
strip~1.
A second numerical problem is related to the
determination of the location of the 1-strings in strip~2. This problem was
previously encountered in \cite{PCA} and arises because $y_{k}^{(2)}$ cannot be
obtained by direct iteration of the auxiliary equation. It is partially solved
here using two different schemes
\begin{enumerate}
\item Inverting certain phases containing $y_{j}^{(1)}-y_{k}^{(2)}$.
\item Inverting the boundary term that contains $y_{k}^{(2)}-\xi$.
\end{enumerate}
The first method, used with success in \cite{OPW}, can be
sensitive to errors in the location of $y_{j}^{(1)}$.
The second method ties $y_{k}^{(2)}$ to $\xi$
so it is generally less sensitive to errors in $y_{j}^{(1)}$. Usually both
schemes converge to the same results but in some regions we find one scheme
converges where the other fails.

\section*{Acknowledgements}

This work is supported by the ARC and INFN. Part of this work was done
during visits of FR to Melbourne and GF and PAP to Bologna and
IPAM, UCLA. We thank these institutes for their support. 
We thank P. Dorey, G. Mussardo, V. Rittenberg and R. Tateo for reading
the manuscript.

\def\baselinestretch{.5}

\end{document}